\documentstyle[twocolumn,prl,aps,epsf,floats]{revtex}

\begin{document}
\draft
\twocolumn[\hsize\textwidth\columnwidth\hsize\csname@twocolumnfalse\endcsname
\title{Spin Reorientations Induced by Morphology Changes in Fe/Ag(001)}
\draft
\author{D.M.~Schaller\cite{dms}, D.E.~B\"urgler, C.M.~Schmidt,
F.~Meisinger, and H.-J.~G\"untherodt}
\address{Institut f\"ur Physik, Universit\"at
Basel, Klingelbergstrasse 82, CH-4056 Basel, Switzerland}
\date{\today}
\maketitle
\begin{abstract}
By means of magneto-optical Kerr effect we observe spin 
reorientations from in-plane to out-of-plane and vice versa upon 
annealing thin Fe films on Ag(001) at increasing temperatures. 
Scanning tunneling microscopy images of the different Fe films 
are used to quantify the surface roughness. 
The observed spin reorientations can be explained with the experimentally 
acquired roughness parameters  by taking into account the effect 
of roughness on both the magnetic dipolar and the magnetocrystalline anisotropy. 
\end{abstract}


]
  
\narrowtext

Ultrathin ferromagnetic films have attracted an enormous interest in 
recent years. Due to the broken symmetry at the surface their 
anisotropies are strongly altered compared to bulk values. In a
phenomenological description 
one can distinguish between bulk and surface contributions to the 
effective anisotropy \cite{fe-jong94-1},
\vspace*{-1ex}
\begin{equation}             
            K^{eff} = K_{v} + \frac{2K_{s}}{d},
\label{e-1}
\end{equation}
where $K^{eff}$ denotes the effective, $K_{v}$ the volume, 
and $K_{s}$ the surface anisotropy. 
In the Fe/Ag(001) system, it has been found that the out-of-plane surface 
anisotropy can dominate the volume anisotropy in the ultrathin range
and align the magnetization perpendicular to the 
surface \cite{fe-jonk86-1,fe-hein87-1,fe-koon87-1,fe-stam87-1,fe-aray88-1}.
Recently, the temperature and thickness dependent 
spin reorientation transition has been further investigated on 
wedge-shaped samples 
\cite{fe-papp92-1,fe-qiu93-1,fe-qiu94-1,fe-berg96-1}. 
In spite of this much better understanding 
of the magnetic phase transition, there is still considerable disagreement in 
literature concerning the exact thickness and temperature where 
the transition occurs, which makes it difficult to compare 
experimental data with theory. As an example, we mention the 
discrepancy between the
results recently obtained by Berger and Hopster \cite{fe-berg96-1} and the 
magnetic phase diagram constructed 
by Qiu~{\it et~al.} in Ref.~\cite{fe-qiu93-1}: For a 4.3 monolayer (ML) thick 
Fe film on Ag(001), Berger and Hopster measure a temperature dependent spin 
reorientation from out-of-plane 
to in-plane at $T {\approx} 220$~K, whereas Qiu {\it et~al.} observe that the 
out-of-plane configuration is stable up to 400~K at this thickness.

The importance of morphology in thin film 
magnetism has been realized from the very beginning of this field and quite some 
discussion arose about the growth mode of Fe on Ag(001)
\cite{fe-hein88-1,fe-egel89-1,fe-li90-1,egel91-1,buer97-1},
but there are hardly any {\em direct} measurements of the influence of 
structure on magnetism.
In this study, we use scanning tunneling microscopy (STM) to get 
direct space information of the morphology of our samples. Direct comparison 
of the structural results with magneto-optical Kerr effect (MOKE)
measurements allows us to observe 
spin reorientations induced solely by morphology changes. 
Morphology dependent spin reorientations have an influence on the 
magnetic phase diagram, an effect that has not been considered in 
literature so far.

\begin{figure*}[bt]    
	\epsffile{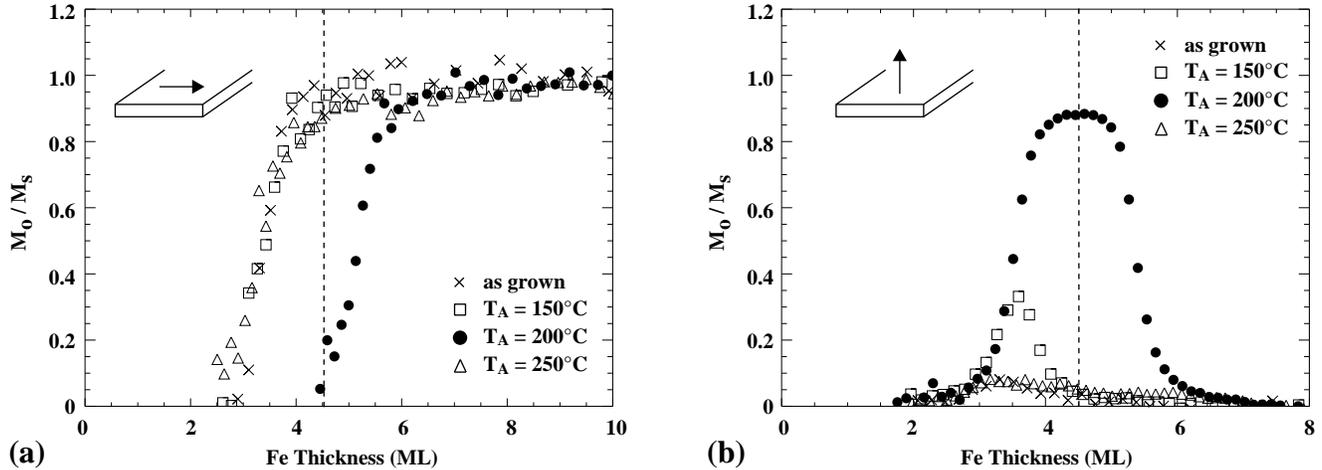} 
	\caption{ Remanent magnetization measured at RT for Fe 
	films grown at RT on Ag(001) and annealed at various 
	temperatures $T_{A}$: 
	(a) longitudinal Kerr signal,
	(b) polar Kerr signal.}
	\label{f-rem}
\end{figure*}

Sample preparation and characterization with the exception of MOKE 
measurements are performed in an ultra-high vacuum system with a 
base pressure of $5 \times 10^{-11}$~mbar which is equipped with
molecular beam epitaxy, STM, low energy electron diffraction 
(LEED), and X-ray 
photoemission electron spectroscopy (XPS). The magnetic measurements 
are performed {\em ex situ} with a MOKE setup that allows the detection of 
longitudinal and polar Kerr effect. All measurements are 
carried out at room temperature (RT). As substrates, we use 150~nm thick Ag films 
which are grown on Fe precovered GaAs(001) wafers. 
After growth at 100$^{\circ}$C and postannealing at 300$^{\circ}$C for one hour,
we get high-quality single crystalline
Ag(001) films. More details about the 
substrate preparation can be found in 
Ref.~\cite{buer96-1}. The Fe films are grown at RT
onto the Ag(001) substrates. The evaporation rate is 
monitored by a quartz thickness monitor and is typically 0.5~ML/min. 
Wedge-shaped Fe films with a slope of 2~ML/mm are grown by linearly moving a 
shutter in front of the 
substrate during deposition. After deposition, the Fe films are 
either investigated as grown or after postannealing for half an hour at 
elevated temperatures.
The temperature accuracy is estimated to be $\pm$10$^{\circ}$C. 
Before the magnetic measurements are performed, the samples are coated 
with 10~nm Ag in order to have two chemically identical interfaces of the Fe 
film and to protect the samples from air exposure.

Longitudinal Kerr effect is used to measure the in-plane component of 
the magnetization. For this purpose, the external magnetic field is 
applied in the plane of the film parallel to a [100] easy axis of the Fe film.
In Fig.~\ref{f-rem}(a), the remanent magnetization 
divided by the saturation magnetization is plotted 
for Fe films directly after growth at RT and after 
postannealing for half an hour at 150$^{\circ}$C, 200$^{\circ}$C, and 
250$^{\circ}$C, respectively. The data points basically lie on two 
different curves: The sample annealed at 200$^{\circ}$C 
shows full remanence down to a thickness of $\sim 5.5$~ML and the 
remanence disappears for smaller thicknesses within a monolayer.
All the other samples show full remanence down to 
$\sim 3.5$~ML and again the remanence fully vanishes for smaller thicknesses 
within a monolayer.

In Fig.~\ref{f-rem}(b), the remanent magnetization is plotted for the different 
samples in polar geometry, which is sensitive to out-of-plane magnetization.
Again, the sample annealed at 200$^{\circ}$C differs substantially 
from all the others: It shows almost full remanence around 4.5~ML, whereas 
all the other samples show no or only very little out-of-plane remanence. 
This means that for Fe films with a thickness of about 4.5~ML, which is 
highlighted in the plots by a dashed vertical line, we observe a spin 
reorientation from in-plane to out-of-plane  upon annealing at 
200$^{\circ}$C. A second spin reorientation back to in-plane is 
observed after annealing at 250$^{\circ}$C.
Because the Curie temperature $T_{c}$ is falling below RT at 2-3 ML 
\cite{fe-qiu93-1}, neither in-plane nor out-of-plane remanent magnetization 
is observed below this thickness.

\begin{figure*}[bt]    
	\epsffile{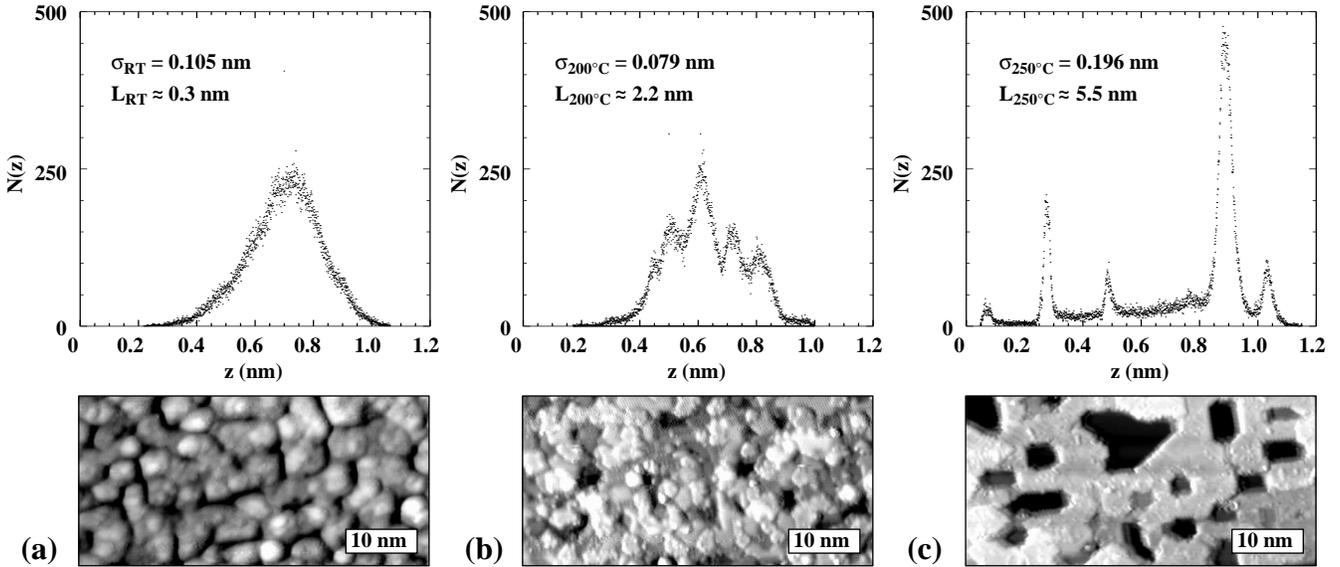} 
	\caption{ STM images (50$\times$25~nm$^2$) and the corresponding 
	histograms of a 4.5~ML thick Fe film grown at RT on Ag(001) and 
	annealed at various temperatures:
	(a) as grown,  
	(b) after postannealing at 200$^{\circ}$C,
	(c) after postannealing at 250$^{\circ}$C. The vertical roughness $\sigma$ 
	and the lateral roughness $L$ extracted from the images are displayed.}
	\label{f-stm}
\end{figure*}

We perform structural analysis of 4.5~ML thick Fe films by LEED 
and STM. By LEED, we observe a sharpening of the spots upon
annealing, in agreement with previous reports \cite{fe-qiu93-1,fe-berg96-1}, 
but it is difficult to get quantitative information about the morphology.
Therefore, we characterize the surfaces also by STM (Fig.~\ref{f-stm}). 
To quantify the morphology, we 
determine the \emph{vertical roughness} $\sigma$ as the average 
deviation from the mean flat 
surface, $\sigma={\langle|\delta z|\rangle}$, and the \emph{lateral roughness} 
$L$ as the average lateral size of terraces.
After growth at RT (Fig.~\ref{f-stm}(a)), 
we observe an irregular arrangement of growth hillocks, in 
accordance with measurements 
of thicker Fe films on Ag(001) \cite{buer97-1}. The vertical roughness 
amounts to $\sigma_{RT}$~=~0.105~nm. The lateral roughness is too 
small to be determined from the STM image, but it must be on the order of the 
Fe lattice constant, hence $L_{RT} \approx 0.3$~nm.
The upper part of the figure displays the histogram of the image, which has a 
Gaussian shape. There cannot be detected any peaks corresponding to Fe terraces.
After annealing at 150$^{\circ}$C, the STM image (not shown) looks very similar 
to Fig.~\ref{f-stm}(a) and also the vertical roughness is only 
slightly reduced to $\sigma_{150^{\circ}{\rm C}}$~=~0.103~nm. Evident morphology 
changes appear after a further annealing at 200$^{\circ}$C: The 
hillocks have transformed into islands with single atomic steps (Fig.~\ref{f-stm}(b)). 
This can be concluded also from the clear peaks in the histogram.
The distances between the histogram peaks correspond to a combination of
Fe steps on different Ag terraces. $\sigma_{200^{\circ}{\rm C}}$ is 
further reduced 
and amounts to 0.079 nm. Because the lateral period is now determined 
by terraces rather than growth hillocks, $L$ can be evaluated by 
calculating the position of the first maximum $R$ in the radial 
height-height correlation function. For morphologies with locally only 
two levels, $L$ corresponds to half the period in the height-height correlation 
function, hence $L \approx R/2$. 
With this statistical method we obtain $L_{200^{\circ}{\rm C}} \approx 2.2$~nm.
Further annealing of the Fe film at 250$^{\circ}$C again
changes the surface drastically (Fig.~\ref{f-stm}(c)): The terraces are 
much larger, $L_{250^{\circ}{\rm C}} \approx 5.5$~nm, and holes 
with a depth of 4-5~ML between them expose the Ag substrate.
These holes cover 25\% of the surface. The appearance of the holes 
increases the vertical roughness to 
$\sigma_{250^{\circ}{\rm C}}$~=~0.196~nm. This value is probably 
underestimated because the STM tip cannot reach the bottom of the 
smaller holes. The histogram exhibits 
much sharper peaks corresponding to Ag (left 3 peaks) 
and Fe step heights.

The morphological changes detected by STM coincide with the chemical 
analysis by XPS measurements: The ratio of the 
Ag 3d peak area to the Fe 2p peak area is constant or even slightly reduced 
after postannealing at 150$^{\circ}$C and 200$^{\circ}$C, in 
correspondence with a reduced surface roughness, and it is increased 
by 15\% after postannealing at 250$^{\circ}$C, which is 
consistent with the appearance of holes exposing the Ag substrate and 
covering 25\% of the area.

The appearance of these holes upon annealing at high enough temperatures 
can be explained by the twice as large surface free 
energy of Fe(001) compared to the value of Ag(001) \cite 
{smit87-1}. If the diffusion length is large enough 
and the coverage is low, the system can minimize its energy by dewetting 
the substrate and forming a Ag surface. The influence of the 
surface free energy on the growth mode of Fe on Ag(001) has been 
discussed in more detail in Ref.\cite{buer97-1}.

Roughness changes both the magnetocrystalline and the magnetic dipolar 
anisotropy. Using N\'{e}el's model \cite{fe-neel54-1}, several authors 
investigated the additional symmetry breaking at steps giving rise to 
a magnetic step anisotropy \cite{fe-brun88-1,fe-albr92-1,fe-chua94-1}. 
The N\'{e}el model fails to predict a 
surface anisotropy for a bcc(001) surface in first-nearest-neighbor 
approximation \cite{strain}. Because 
second-nearest-neighbors 
are only 15\% more distant than first-nearest-neighbors in a 
bcc crystal, they must not be neglected and the N\'{e}el 
model predicts indeed an out-of-plane surface anisotropy in this approximation. 
Second-nearest-neighbors have simple cubic coordination.
For this symmetry, Bruno calculated 
in Ref.~\cite{fe-brun88-1} a decrease of the out-of-plane surface 
anisotropy by 50\% 
for each step atom. This reduction needs to be multiplied by the 
percentage of step atoms at the surface, which can be counted to 
be $4\sigma / L$ \cite{fe-brun88-1}. 
This results in a decrease of the surface anisotropy by 
$\Delta K_{s}/{K_{s}} = -2\sigma/L$.

Using the experimentally acquired roughness parameters, we calculate 
a decrease of $K_{s}$ by more than 50\% for the sample prepared at RT, 
whereas after annealing at 200$^{\circ}$C and 250$^{\circ}$C, the 
reduction amounts only to 7.2\% and 7.1\%, respectively. Consequently, 
the reduction of $K_{s}$ induced by roughness can explain the spin reorientation 
from in-plane to out-of-plane upon annealing the Fe film at 
200$^{\circ}$C, but it cannot account for the second spin reorientation back to 
in-plane after further annealing at 250$^{\circ}$C. This second spin 
reorientation is caused by the reduction of the interface area
by 25\% due to the holes which results in a reduction of $K_{s}$ 
by the same amount. Additionally, we have to take into account the 
effect of roughness on the magnetic dipolar anisotropy.
Bruno calculated in Ref.~\cite{fe-brun88-2} that roughness also gives rise 
to a purely dipolar surface anisotropy. 
If the vertical roughness $\sigma$ is dependent on thickness, 
this anisotropy reduces the absolute value of $K_{v}$. A strong thickness 
dependence of the vertical 
roughness is expected for the sample annealed at 
$250^{\circ}$C, because the hole depth 
and thereby also $\sigma$ linearly increase with thickness.

\begin{figure}[tb]
    \epsffile{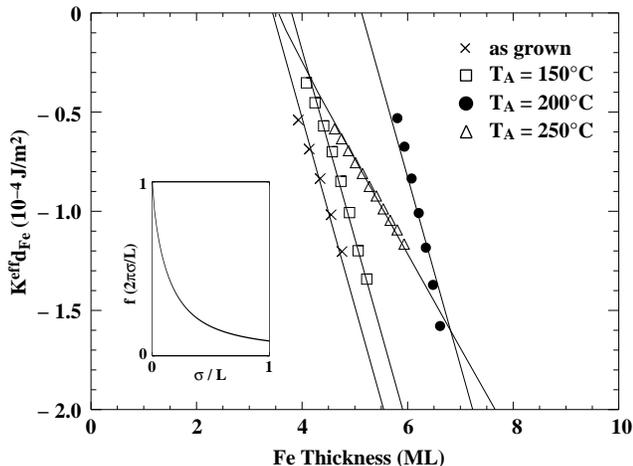}
    \caption{The product of the effective anisotropy and the Fe 
    thickness is plotted as a function of the Fe thickness for Fe 
    films grown at RT and annealed at various 
    temperatures $T_{A}$. The solid lines are a simultaneous fit of the data to 
    Eq.~(\ref{e-3}). The inset shows the graph of the function $f$ used 
    in Eq.~(\ref{e-3}).}
\label{f-hs}
\end{figure}

Summarizing the effect of roughness on magnetic anisotropy, we can 
rewrite Eq.~(\ref{e-1}), taking into account the effect of roughness at 
one of the interfaces:
\vspace*{-1ex}
\begin{eqnarray}             
            K^{eff} = &-& \frac{1}{2} \mu_{0} M_{s}^{2} + 
            \frac{2K_{s}}{d} \Bigl(1 - 2\alpha\frac{\sigma}{L}\Bigr) \nonumber\\
            &+& \frac{3}{8} \mu_{0} M_{s}^{2} \frac {\sigma} {d}
            \Bigl( 1 - f \Bigl(2 \pi \frac {\sigma} {L}\Bigr)\Bigr) .
\label{e-3}
\end{eqnarray}
The last term is the roughness induced dipolar surface anisotropy 
calculated in Ref.~\cite{fe-brun88-2}. The graph of the function $f$ 
which is given explicitly in Ref.~\cite{fe-brun88-2} is plotted in 
the inset of Fig.~\ref{f-hs}. The parameter $\alpha$ denotes the 
reduction of the surface anisotropy per step atom,
which is predicted to be 50\% in the N\'{e}el model.

To quantify our observations, we determine the effective anisotropy of the 
different samples. Elementary electromagnetic considerations 
show that the area enclosed between polar and longitudinal hysteresis loops
is proportional to $K^{eff}$ \cite{fe-jong94-1}. In the 
simple case where we measure square longitudinal loops, the polar
saturation field, the so called anisotropy field $H_{A}$, is 
proportional to the effective anisotropy and $K^{eff} = 
-\mu_{0}H_{A}M_{s}/2$, where $M_{s}$ denotes the saturation 
magnetization. 

In Fig.~\ref{f-hs}, the product of the effective 
anisotropy and the Fe thickness is plotted as a function of the Fe
thickness for the differently prepared samples. The solid lines are 
the result of a simultaneous fit of the data to Eq.~(\ref{e-3}). The 
roughness parameters determined from the STM images are used and the
vertical roughness of the sample annealed at $250^{\circ}$C is assumed 
to be linearly dependent on thickness, $\sigma_{250^{\circ}{\rm C}}(d) = 
\sigma_{0}d$, in agreement with the linear increase of the hole depth 
with thickness. The decrease of $K_{s}$ by 25\% for the sample 
annealed at $250^{\circ}$C is taken into account.

The best fit is achieved with the following parameters: $\alpha = 64\%$, 
$M_{s} = 1.03$~MJ/m$^{3}$, $K_{s} = 0.27$~mJ/m$^{2}$, 
$\sigma_{250^{\circ}{\rm C}}(d)$~=~0.81$d$, and 
$L_{150^{\circ}{\rm C}} = 1.16 L_{RT}$. 
The reduction factor $\alpha$ is indeed fitted very close to 50\% 
which is what is expected from the N\'{e}el model in second-nearest-neighbor 
approximation. The other parameters are in 
reasonable agreement with the expected values: 
$L_{150^{\circ}{\rm C}}$ is only slightly larger than $L_{RT}$, and 
$\sigma_{250^{\circ}{\rm C}}$ has a thickness dependence close to 
$\sigma(d) = 0.5d$ valid for a flat film perturbed by holes making up 
25\% of the area.
The values of $M_{s}$ and $K_{s}$ are fitted rather small (see, e.g., 
Ref.~\cite{fe-jong94-1}), which is 
probably due to the fact that the roughness of the bottom interface is 
not taken into account in this analysis. The assumption of equal 
roughness for the bottom interfaces of all samples increases $M_{s}$ and $K_{s}$ 
without further affecting the above analysis.

In conclusion, we have shown by a combined STM and MOKE study that 
the magnetic anisotropy of thin Fe films can considerably be 
altered by morphology. Spin reorientations induced solely by morphology changes 
are observed which can explain some of the discrepancies reported in 
literature concerning the magnetic phase diagram of Fe/Ag(001). The results 
are explained by a change of both the magnetic dipolar and the
magnetocrystalline anisotropy due to roughness. The change of the 
magnetocrystalline anisotropy cannot be predicted within the N\'{e}el 
model in first-nearest-neighbor approximation, even if strain 
is taken into account. In second-nearest-neighbor approximation, which 
may be important in a bcc crystal with small strain, the N\'{e}el 
model predicts a reduction of the out-of-plane surface anisotropy 
which is in close agreement with our experiments.

We would like to thank P. Bruno and R.~C. O'Handley for helpful discussions.
Financial support from the Swiss National Science Foundation and the 
Swiss {\em Kommission f\"ur Technologietransfer und Innovation} 
is gratefully acknowledged.


\begin{references}
\bibitem{strain} 
The symmetry breaking due to strain gives rise to an out-of-plane
surface anisotropy proportional to the misfit \cite{fe-chua94-1}, 
which is very small in the well lattice-matched Fe/Ag(001) system. 
Strain effects do not change our conclusions and are therefore neglected.
\end{references}


\end{document}